# A computational study for tomographic imaging of temperature, water and carbon dioxide concentration in combustion gases using a single tunable diode laser

**Satwik Choudhury[1] and Sandip Pal[1,2] ***

[1]*Homi Bhabha National Institute, 2nd floor, BARC Training School Complex, Anushaktinagar, Mumbai, India - 400 094*
[2]*Variable Energy Cyclotron Centre, 1/AF, Bidhan Nagar, Kolkata, India - 700064*
[*]*sandip@vecc.gov.in*

**Abstract:** There is a present-day need of optimization of the combustion processes and catalytic efficiency for minimization of emission of pollutants, which can be possible by analyzing the temperature and combustion products distribution in the combustion processes internally. This paper discusses the feasibility of simultaneous 2-D tomographic imaging of temperature and concentrations of $CO_2$ and $H_2O$ by scanning a single narrowband laser. The choice of spectroscopic lines of $H_2O$ for two-tone temperature measurement is discussed around the wavelength range of 2000 nm. This region is selected as both $H_2O$ and $CO_2$ spectral lines have sufficient intensity levels. The pair of wavenumbers (5005.53 and 5003.3 $cm^{-1}$) is found the best of all for temperature and $H_2O$ concentration distribution and 5004.36 $cm^{-1}$ is chosen for $CO_2$. To establish the efficacy of different reconstruction algorithms, the phantoms of bimodal and concentric temperature distribution with uniform concentration were tried with four reconstruction techniques. Impressive reconstructions were achieved for distribution of temperature and concentrations of $H_2O$ and $CO_2$ for same types of phantoms using Filtered Landweber and Tikhonov regularization methods for 5005.53 and 5003.3 $cm^{-1}$. Initially the imaging was tried using both fanbeam and discrete irregular beam array with 100 and 31 beams respectively. Instead of less number of beams, the later one also shows very promising result.

## 1. Introduction:

The combustion structure and the combustor efficiency of Internal combustion (IC) engines, burners, gas turbines and so on are required to be optimized. Thus, real time 2-D temperature and concentration distributions are important factor in $NO_x$, total hydrocarbon (THC) and particulate matter (PM) emissions and can be used to analyze catalytic efficiency for diesel and gasoline engines and combustion oscillations of gas turbines [1]. Tunable diode laser absorption spectroscopy (TDLAS) has become indispensable tool since past few decades for combustion studies and flow diagnostics as it is capable of measuring various gas parameters like temperature, pressure, concentration etc. with reliable sensitivity even in harsh environments [2-7]. The advantages of absorption spectroscopy are specificity to chemical species, versatility of measurements of multiple parameters, and simplicity of implementation etc. [8]. The traditional line-of-sight (LoS) measurements can only provide the path averaged gas-parameters, while the parameters are most likely to exhibit non-uniformity along the path. In non-uniform flow fields to resolve distinct spatial absorption features, successful tomographic reconstruction is possible by using LoS measurements along multiple directions, either by parallel or fan-beam arrangement [9-10]. Angular displacements were achieved by rotating the combination of source-detector combination around the object, achieving good spatial resolution but sacrificing temporal resolution. In fast changing turbulent flows, the former method is not applicable and simultaneous detection of transmitted light in multiple array of photodetectors, i.e., TDLAS at different multiple paths can help to perform chemical-species tomography, inspired by X-ray Computed Tomography in medical fields. Multiple path launching of laser is possible either by using open-path splitter [11,12,13] or by using fiber splitter [9,14].

Wang and his co-researchers developed a tomographic system capable of reconstructing 2D distribution of $H_2O$ concentration and temperature using algebraic reconstruction technique utilizing the spectral lines 7153.748 and 7154.354 $cm^{-1}$ [14]. Liu et al obtained temperature and concentration fields of $H_2O$ in a $CH_4$ flat flame [11] and swirl injector [15] using two distributed feedback lasers at 7185.6 $cm^{-1}$ and 7446.36 $cm^{-1}$ based on the regularized

Abel Onion Peeling inversion transform method and three-point method, respectively. Wang demonstrated a tomographic sensor capable of reconstructing $NH_3$ concentration and temperature (using the transitions at 6544.4 and $6543.9 cm^{-1}$) within very short time ($\sim 100 mS$) utilizing cylindrical retro-reflectors [12]. Reconstruction of carbon-di-oxide temperature and concentration profile (using v3 bandhead region of $CO_2$ at 4.2 $\mu m$) in concentric methane co-flow diffusion flame using TDLAS was reported by Zhang et al [16]. Multiple species tomography was also performed by Kamimoto's group [1] where they successfully reconstructed $NH_3$ and $CH_4$ distributions using the transitions 6612.79 $cm^{-1}$ ($NH_3$) and 6114.94 $cm^{-1}$ ($CH_4$) for the concentration measurements. Ma et al. built a hyper-spectral tomography system for detecting 2-D temperature and $H_2O$ concentration in the exhaust flue gas of a J85 engine [17]. They have employed a time-division-multiplexed combination of three Fourier-domain mode-locked (FDML) lasers and the system shows great capability and efficiency except for the high cost. Grauer et. al. [18] demonstrated linear hyperspectral absorption spectroscopy by reconstructing three physical parameters, i.e., concentration, temperature, pressure, simultaneously, from experimental absorbance data of a number of spectral lines with the wavelength scan range of a single laser. All the above-mentioned works are summarised in Table 1.

| Measurement parameters | Wavenumber #1 $cm^{-1}$ | Wavenumber #2 $cm^{-1}$ | Chemical Species | Reference |
|---|---|---|---|---|
| Temperature & Concentration | 7153.748 | 7154.354 | $H_2O$ | Wang et al 2015 [14] |
| Temperature & Concentration | 6544.4 | 6543.9 | $NH_3$ | Wang et al 2010 [12] |
| Temperature & Concentration | 2397.277 | Onion Peeling and Fourier analysis | $CO_2$ | Zhang et al 2018 [16] |
| Temperature & Concentration | 7185.6 | 7446.36 | $H_2O$ | Liu at al 2015, 2017 [11,14] |
| Temperature & Concentration | 7203.910<br>7202.919<br>7202.255 ($H_2O$) | 6612.794 ($NH_3$)<br>6114.936 ($CH_4$) | $H_2O$, $NH_3$, $CH_4$ | Deguchi et al 2015 [1] |
| Temperature, Pressure & Concentration | Range in between 7404.5 and 7405 | | $H_2O$ | Grauer et. al. 2019 [18] Hyperspectral absorption |

Table 1: Classification of different earlier studies for multi-parameter tomographic imaging

The spatial resolution of reconstruction can be increased by increasing the number of multiple paths. Space-limitation restricts the number of source and detector array pairs around the object for simultaneous detection resulting a fewer projection data in contrast to medical scanners (where either the object or the source and detector array pair can be rotated), sacrificing spatial resolution. Hence, solving the inverse problem computationally is more challenging. Instead of filtered back-projection method, one usually employs here various iterative reconstruction techniques, like Algebraic Reconstruction Technique (ART), Landweber method, Simultaneous Iterative Reconstruction method (SIRT) etc. Wood and Ozanyan [19] performed numerical study on reconstruction of $H_2O$ concentration and temperature distribution using Landweber iteration. Li and Weng [20] discussed a modified version of ART named MAART, with an auto-adjustment relaxation parameter and made a comparative study of the reconstruction quality with those of SIRT and ART through numerical simulation. Grauer et al. [18] used Tikhonov and Exponential prior-based technique for implementing hyperspectral absorption tomography.

In researches [14,17,18], simultaneous temperature and concentration distribution of a single species were attainable using the scan of a single laser diode, but researchers employed multiple laser diodes to measure the temperature and concentration distribution of chemical species more than one [17]. In the present work, numerical investigations have been performed to establish the possibility of reconstruction of the temperature and concentration distributions of two chemical species, i.e. $CO_2$ and $H_2O$, which are the major combustion products in most of the combustion systems, by scanning driving current of a single narrowband laser diode. This novel approach will greatly reduce and simplify the hardware requirement for detection of possible absorption feature. To assess the quality of reconstruction, several well-known algorithms are utilized to have an idea which can provide the best performance over the others to reconstruct the required distributions.

## 2. Fundamental Equations of absorption and temperature detection:

When an incident collimated monochromatic laser light of intensity, $I_0(\nu)$ and wavenumber, $\nu(cm^{-1})$ passes through a medium with a pathlength of $L(cm)$. The transmitted intensity of radiation, $I_t(\nu)$ through a medium is described by well-known Beer-Lambert law, which can be written as,

$$\ln \frac{I_t(\nu)}{I_0(\nu)} = -\beta_\nu \quad (1)$$

where

$$\beta_\nu = \int_0^L P(l).X(l).S_{\nu i}[T(l)].\phi(\nu).dl \quad (2)$$

where, $P(l)$ and $T(l)$ are local total pressure and temperature respectively, $X(l)$ is local concentration of absorbing species, $S[T(l)]$ is line strength function, and $\phi(\nu)$ is area-normalized line shape function. The variation of line-strength $S(T)$ is a function of local temperature T [21] can be expressed like,

$$S(T) = S(T_0) \frac{Q(T_0)}{Q(T)} \frac{T_0}{T} \exp\left[-\frac{hcE''}{k_B}\left(\frac{1}{T} - \frac{1}{T_0}\right)\right] \times \frac{1-\exp\left(-\frac{hc\nu_0}{k_B T}\right)}{1-\exp\left(-\frac{hc\nu_0}{k_B T_0}\right)} \quad (3)$$

where, $Q(T)$ is the molecular partition-function, $T_0$ and $p_0$ is a reference temperature and pressure, $\nu_0$ is the line center wavenumber, $E''$ is lower energy-state of transition, $h$ is Planck's constant, $c$ is speed of light in vacuum and $k_B$ is Boltzmann's constant.

The line-shape function $\phi(\nu)$ can be represented by Voigt profile, which can be expressed as a convolution of Lorentz and Doppler broadening mechanism.[21]

$$\phi_V(\nu) = \frac{2}{\omega_G}.\sqrt{\frac{\ln 2}{\pi}}.\frac{a}{\pi}.\int_{-\infty}^{\infty} \frac{\exp(-y^2)}{a^2+(\omega-y)^2} dy \quad (4)$$

where, parameter $\quad a = \sqrt{\ln 2}.\frac{\omega_L}{\omega_G} \quad (5)$

variable $\quad \omega = \frac{2.\sqrt{\ln 2}}{\omega_G}.(\nu - \nu_i) \quad (6)$

Integral variable $\quad y = \frac{2\sqrt{\ln 2}}{\omega_G} u \quad (7)$

$\omega_G$ = Gaussian Full Width Half Maximum (FWHM) $= \frac{2\nu_0}{c}.\sqrt{\frac{2N_A kT \ln 2}{M}} \quad (8)$

$\omega_L$ = Lorentz FWHM $= 2\left(\frac{T_0}{T}\right)^{n_{ai}}.\left[\gamma_{air}(p_0,T_0).(p-p_{self}) + \gamma_{self}(p_0,T_0).(p_{self})\right] \quad (9)$

M= Molecular Weight and $N_A$ = Avogadro Number = 6.023 x $10^{23}$ mol$^{-1}$

$\gamma_{air}$ and $\gamma_{self}$ = The air-broadened and self-broadened half width at half maximum (HWHM) (cm$^{-1}$/atm), respectively at $T_0$=296K and reference pressure $p_0$=1atm

$p_{self}$ = partial pressure of the self

$n_{air}$ = The exponent of the temperature dependence of the air and self-broadened half width

$\delta_{air}$ = The pressure shift (cm$^{-1}$/atm) at $T_0$=296 K and $p_0$=1atm of the line position with respect to the vacuum transition wavenumber, $\nu_0$

It is observed that the ratio of the absorption coefficients for total integrated absorption of two pre-selected transitions at grid '*j*' from equation (3) can be expressed using analytical equation [21] as,

$$R_j = \frac{\beta_{1j}}{\beta_{2j}} = \frac{S_1(T_0)}{S_2(T_0)} exp\left[-\frac{hc(E''_1 - E''_2)}{k_B}\left(\frac{1}{T} - \frac{1}{T_0}\right)\right] \tag{10}$$

Hence, we can define absolute temperature sensitivity [21] as,

$$\frac{dR_j}{dT} = \frac{hc(E''_1 - E''_2)}{k_B T^2} \frac{S_1(T_0)}{S_2(T_0)} exp\left[-\frac{hc(E''_1 - E''_2)}{k_B}\left(\frac{1}{T} - \frac{1}{T_0}\right)\right] \tag{11}$$

Also a relative sensitivity [22,23] can be defined as,

$$\xi = \frac{dR_j/R_j}{dT_j/T_j} = \frac{hc}{k_B} \frac{(E''_1 - E''_2)}{T_j} \tag{12}$$

Therefore, from equation (12) it is clear that to achieve a good sensitivity in the reconstruction process, $\Delta E''$ for the two transitions should be sufficiently large.

For tomographic purpose, in order to get 2-D distributions of temperature and gas concentrations, the region of interest (ROI) is discretized into "*J*" number of square grids. It is assumed that within each grid region temperature, pressure and species concentration are uniform. 2-D grid centre coordinate is designated as variable (*m,n*). Except ROI, other region is masked.

When a laser beam "*i*" passes through the ROI, the integrated absorbance of transition $\nu_i$ over path can be written in a discretized form as,

$$A_{\nu i} = \sum_{j=1}^{J} L_{ij} \alpha_{\nu,j} = \sum_{j=1}^{J} L_{ij} \cdot [P.X.S(T)]_{\nu,j} \tag{13}$$

where, $A_{\nu,i}$ is the total absorbance integrated over wavelength and integrated along *i*-th ray, $L_{ij}$ is the length of the *i*-th ray in *j*-th grid creating sensitivity matrix $L$ and $\alpha_{\nu,j}$ is the integrated absorption co-efficient in *j*-th grid for unit length.

In matrix form, $\qquad\qquad\qquad\qquad \boldsymbol{L.\alpha_\nu = A_\nu} \tag{14}$

where, $\boldsymbol{L} = \begin{bmatrix} L_{11} & L_{12} & \cdots & L_{1N} \\ L_{21} & L_{22} & \cdots & L_{2N} \\ \vdots & \vdots & \ddots & \vdots \\ L_{M1} & L_{M2} & \cdots & L_{MN} \end{bmatrix}, \quad \boldsymbol{\alpha_\nu} = \begin{bmatrix} \alpha_{\nu 1} \\ \alpha_{\nu 2} \\ \vdots \\ \alpha_{\nu N} \end{bmatrix}$ and $\boldsymbol{A_\nu} = \begin{bmatrix} A_{\nu 1} \\ A_{\nu 2} \\ \vdots \\ A_{\nu N} \end{bmatrix}$

Now utilizing suitable reconstruction algorithm, $\alpha_{\nu_1,j}$ and $\alpha_{\nu_2,j}$ for $j^{th}$ grid can be satisfactorily reconstructed, for two pre-selected transitions, and their ratio, $R_j$ is independent of concentration. $R_j$ follows a best-fitted empirical relation with $T_j$ similar to equation (13), which has been generated from HITRAN 2012 database [24].

$$T_j = \sum_{k=0}^{5} c_k \cdot R_j^k \tag{15}$$

If $T_j$ is known, integrated (over wavelength) absorption coefficient distribution for each grid for maximum concentration ($\chi^{max}$), $\alpha_{\nu_1,j}^{max}$ can be calculated from the sensitivity matrix for a particular $T_j$. From the actual integrated absorption, $\alpha_{\nu_1,j}$, already computed using reconstruction algorithm, $\chi_j$ can be calculated as,

$$\chi_j = \frac{\alpha_{\nu_1,j}}{\alpha_{\nu_1,j}^{max}} \cdot \chi^{max} \qquad (16)$$

In order to evaluate the concentration distribution of $CO_2$, the reconstruction algorithm can be applied to find the integrated absorption in each grid. Just like $H_2O$, integrated absorption distribution for each grid for maximum concentration can be calculated from the sensitivity matrix utilizing the temperature distribution. For $CO_2$ also, equation (16) is utilized to evaluate the concentration distribution of $CO_2$.

### 3. Selection of absorption lines:

For evaluation of the volumetric concentration of different combustion products, it is necessary to consider full combustion of simple hydrocarbons ($C_nH_{2n+2}$) like methane, ethane, propane, and n-butane for n = 1,2,3 and 4, respectively. Assuming atmospheric air contains oxygen and nitrogen in the volumetric ratio of 1:4, the chemical reaction for can be represented as

$$2C_nH_{2n+2} + (3n+1)(O_2 + 4N_2) = 2n.CO_2 + 2(n+1).H_2O + 4(3n+1).N_2 \qquad (18)$$

For propane, the volumetric ratio of major combustion products like $CO_2$, $H_2O$ and $N_2$ is 6:8:40 ~ their volumetric percentages are 11%, 15% and 74%.

Since CO, $CO_2$ and $H_2O$ are major combustion products in many combustion systems, measurement of their concentration can provide us information regarding combustion efficiency. Also measurements of CO and $CO_2$ can yield estimation about fuel input in combustors and production of CO is an indicator of completeness of combustion. Lean combustion technology is adopted for realization of cleaner and more environmentally friendly power generation. Lean combustion technology means low fuel/air equivalence ratios and is characterized by low flame temperature. Hence, for characterization of burning flame temperature distribution as well as water and $CO_2$ distribution inside and outside the flame is necessary to understand the burning characteristic non-invasively.

Table 2: Properties of suitable absorption lines for temperature and concentration distribution [24]

| Wavenumber $\nu$ (cm$^{-1}$) | | $S$ (cm$^{-1}$/ (molecule.cm$^{-2}$)) | $E''$ (cm$^{-1}$) | $\gamma_{air}$ cm$^{-1}$.atm$^{-1}$ | $\gamma_{self}$ cm$^{-1}$.atm$^{-1}$ | $n_{air}$ | $\delta_{air}$ cm$^{-1}$.atm$^{-1}$ |
|---|---|---|---|---|---|---|---|
| $H_2O$ | 5290.26 cm$^{-1}$ | $1.574 \times 10^{-20}$ | 79.4964 | 0.089 | 0.539 | 0.72 | -0.006 |
| | 5000.22 cm$^{-1}$ | $2.525 \times 10^{-24}$ | 2358.2996 | 0.011 | 0.161 | 0.37 | -0.018 |
| | 5000.06 cm$^{-1}$ | $3.201 \times 10^{-26}$ | 3391.1269 | 0.0185 | 0.176 | 0.41 | -0.01511 |
| | 5005.53 cm$^{-1}$ | $6.426 \times 10^{-25}$ | 2631.2671 | 0.0134 | 0.152 | 0.37 | -0.01567 |
| | 5003.30 cm$^{-1}$ | $4.380 \times 10^{-25}$ | 2358.3005 | 0.0094 | 0.152 | 0.37 | -0.01987 |
| $CO_2$ | 5001.49 cm$^{-1}$ | $3.902 \times 10^{-22}$ | 519.5350 | 0.0683 | 0.084 | 0.76 | -0.005913 |
| | 5002.48 cm$^{-1}$ | $3.100 \times 10^{-22}$ | 578.0116 | 0.0681 | 0.082 | 0.76 | -0.00594 |
| | 5004.36 cm$^{-1}$ | $1.855 \times 10^{-22}$ | 704.3005 | 0.0679 | 0.079 | 0.75 | -0.005983 |

For monitoring temperature, $H_2O$ is the target absorbing species and optimal spectral line pair is to be chosen for designing two-tone absorption thermometry. From $H_2O$ absorption spectra, the pair of lines, having the criteria of sufficiently large $\Delta E''$ (equation (12)) and similar lineshape function, is considered as per table 2. Figure 1 shows the spectra of $H_2O$ and $CO_2$ in our region of interest using HITRAN [24] database at 296K and 1 atm pressure.

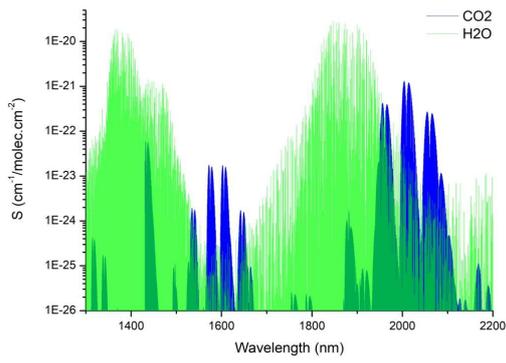
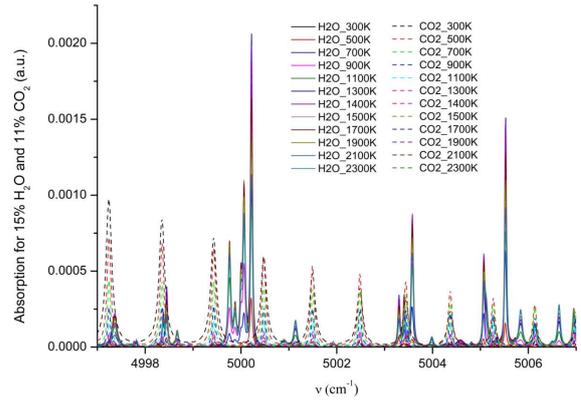

Figure 1: Spectral line intensity for $CO_2$ and $H_2O$ at 296K from HITRAN 2012 [24]

Figure 2: Absorption spectrum for 15% $H_2O$ and 11% $CO_2$ for 0.1 cm pathlength for temperatures from 300K to 2300K

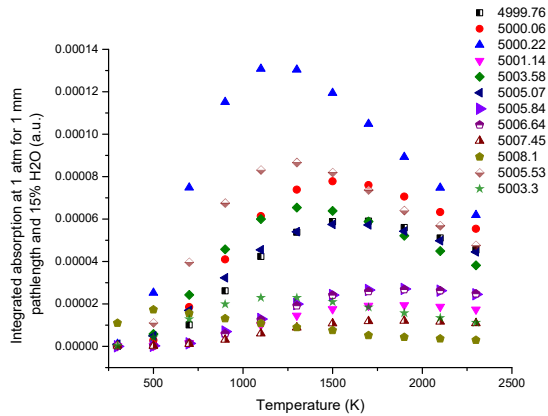
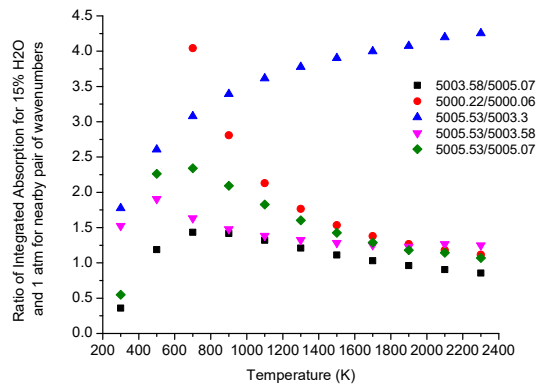

Figure 3: Integrated absorption data for 0.1 cm pathlength with 15% $H_2O$ at 1 atm. pressure for temperature range from 300 to 2300K for different wavelengths

Figure 4a: Distribution of ratio of integrated absorption for several nearby lines, among which two pairs 5005.53/5003.3 $cm^{-1}$ and 5000.22/5000.06 $cm^{-1}$ are found best among all

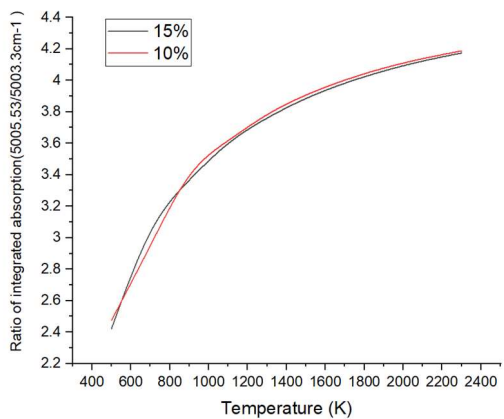
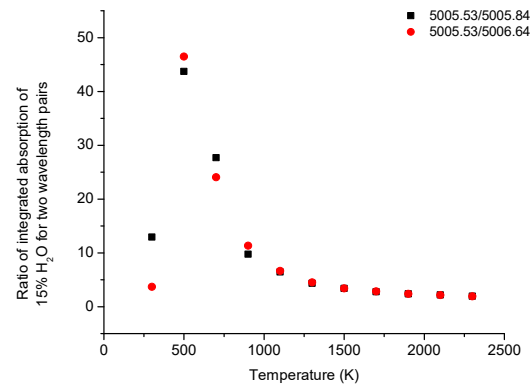

Figure 4b: Variation of ratio of integrated absorption with temperature at two different concentration of $H_2O$

Figure 5: Distribution of ratio of integrated absorption for two pair of wavenumbers 5005.53 with respect to 5005.84 and 5006.64 $cm^{-1}$

Zhou et al [25] discussed wavelength selection criteria for range of wavelength for range of wavelength below 1650 nm. The observations for selection of spectral lines for H2O and CO2 are discussed in detail as below:

#1 To fulfil the criteria of scanning by a single laser source, the two lines for $H_2O$ and one line for $CO_2$ should be close enough in the driving current scanning range at a fixed temperature but without interfering each other. 1300-1650 nm region is chosen in earlier experiments for detecting temperature by using two spectral lines of $H_2O$ because at this region telecommunication lasers and fibers with attractive features at low cost are available. In this region, the best lines were chosen for detection of temperature for tomography by different researchers [22]. As simultaneous detection of $CO_2$ from the same laser source is our target, it is clear from figure 1 that comparable line strengths as of $H_2O$ having considerable detection limit are only available above 1950 nm. Hence, more investigation was done to find out suitable spectral lines in between 1950 nm and 2000 nm.

#2 The distribution of absorption lines for 15% $H_2O$ and 11% $CO_2$ for 0.1 cm pathlength at 1 atm pressure for temperature range from 300K to 2300K are shown in figure 2 for wavenumber scan range from 4997.5 $cm^{-1}$ to 5006.5 $cm^{-1}$. The absorption levels for most of the lines of $H_2O$ are increasing with temperature with slight variation in peak positions and slopes, whereas that for $CO_2$ is decreasing. Few absorption lines of $H_2O$ and $CO_2$ are overlapping at the wing region and influencing each other. Because of the influence of spectral lines of other combustion products, integrated absorption creates erroneous result. Therefore, the pair of lines should be such that overlapping zones of the nearby lines are at the far end of the wing and while calculating the integrated absorption, integration from first minimum of one side to first minimum to the opposite side of the spectral line are considered.

#3 Both the lines of the selected pair should have sufficient absorption strength over the selected temperature range from 500K to 2300K. For tomography, region of interest of diameter 50 mm is divided into square grid of dimension of 1 mm. In order to attain spatial resolution of 1 mm, the absorption of particular wavelength for one grid length should be more than the detection limit. Figure 3 shows the minimum integrated absorption of the order of $10^{-5}$ for 1 mm pathlength, the detection limit should be $10^{-6}$ of the transmitted light considering SNR=10.

#4 The absorption ratio of two wavelengths should vary monotonically with temperature. Figure 4a shows that the variation of the ratio of integrated-absorptions of wavenumber pairs 5003.58 and 5005.53 $cm^{-1}$ are small resulting in low sensitivity, whereas that for wavenumber pairs 5005.53 and 5005.07 $cm^{-1}$ are not monotonically varying. The same for the wavenumber pairs 5005.53 and 5003.3 $cm^{-1}$ varies with temperature monotonically, increasing from 1.6 to 4.26 for temperature from 300 K to 2300 K. The sensitivity of the curve is around 0.0016/K, for which the temperature variation can be easily sensed. Error analysis shows that if the maximum uncertainty in ratio measurement is $2 \times 10^{-6}$ considering relative error in absorption as $1 \times 10^{-6}$. The ratio for 5000.22 and 5000.06 $cm^{-1}$ varies from 8.79 to 1.12 for 500 to 2300 K having sensitivity as 0.0019/K but useful above 700K as the measurement for 5000.06 $cm^{-1}$ below 700K is very low introducing large uncertainty. Hence, suitable pairs are 5005.53 & 5003.3 $cm^{-1}$ and 5000.22 & 5000.06 $cm^{-1}$.

Figure 4b shows the result of variation of ratio for the wavenumber pairs 5005.53 and 5003.3 $cm^{-1}$ for changing the concentration from 15% to 10% $H_2O$. The difference between the two curves is less than 0.5% in ratio, which in turn introduces error of around 30K in 2000K range. Figure 5 shows the variation of ratio of integrated absorptions of 5005.53 $cm^{-1}$ with 5005.84 and another with 5006.64 $cm^{-1}$ vs. temperature. In both the cases at 500K the ratio value is around 50, which reduces to less than 5 at 1300K. From figure 3 it is clear that the integrated absorption values for the later ones are quite low below 1300K, introducing large uncertainty in the measurement result.

#5 The surrounding environment is chosen a 2% moisture at 500 K. In order to alleviate its effect, one solution is to flush the surrounding with dry nitrogen, which is very cumbersome process. Other solution is to choose the spectral lines in such a way that the line strength is much stronger at temperature 700 K – 2300 K than the absorption near the ambient temperature. It is clearly found that at 5005.53 $cm^{-1}$ and 5003.3 $cm^{-1}$ the integrated absorption values for 2% $H_2O$ and 1 mm pathlength at 500K are $1.5 \times 10^{-6}$ and $5.65 \times 10^{-7}$, i.e. both are of the order of the detection limit.

Hence, first of all, a special case is considered in which only temperature is varying but the concentrations of the absorbing species ($H_2O$ and $CO_2$) are constant around the region. Because of the non-linear relationship in between integrated absorbance and the temperature, its effect on the temperature tomography is critical. The more general case, when both temperature and concentrations of absorbing molecules vary over the region of combustion, will be considered later. It can be shown that three distributions (temperature, concentration of $H_2O$ and $CO_2$) can be reconstructed simultaneously, only by scanning one laser for a fixed temperature across the possible spectral lines.

From the above discussion, it is found that the pair #1, e.g. 5005.53 and 5003.3 $cm^{-1}$, and pair #2, e.g. 5000.22 and 5000.06 $cm^{-1}$ are found best suited for temperature and water concentration measurement. The $CO_2$ lines at 5001.49, 5002.47, and 5004.36 $cm^{-1}$ are free from any interference; hence, can be utilised for $CO_2$ concentration measurement.

Irrespective of the above two pair of lines, another prospective line ν3 of $H_2O$ is at 5290.264 $cm^{-1}$ to create two new pairs of $H_2O$ lines with 5005.53 and 5000.22 $cm^{-1}$. The corresponding values of $E''$ and S (at 296 K) are tabulated in Table 2 (from HITRAN 2012 database [24]). Figure 6 shows the spectral lines in the range 5290.264 $cm^{-1}$ for 15% $H_2O$ at 1 atm. pressure and for the temperature range in between 700K to 2300K. The plot figure 7(a) shows that the integrated absorption with temperature decreases for 5290.264 $cm^{-1}$, whereas the same increases for wavenumbers 5005.53 and 5000.22 $cm^{-1}$. The ratio of integrated absorption for the pairs 5005.53/5290.264 $cm^{-1}$ and 5000.22/5290.264 $cm^{-1}$ increases from 0 to 4.5 and 5.5, respectively as shown in figure 7(b). It is evident that the sensitivities for these pairs are around 2.5E-03/K and 3.0E-03/K, respectively, almost 1.5 times the pair #1 and #2, discussed in earlier section. Therefore, the pairs, discussed in this section, provide a better sensitivity than the other two pairs due to having a much larger $\Delta E''$ as evident from Table 2. The main drawback of the new pairs is that they cannot be scanned by using a single narrow-band laser. The combination of wavenumbers 5005.53/5290.264 $cm^{-1}$ is treated as pair #3 in onward discussion.

## 4.    Hardware Arrangement:

To perform tomography, usually a direct-absorption technique is employed. Two options are tried one with fan-beam arrangement and another comprising of few discrete rays with uneven distribution.

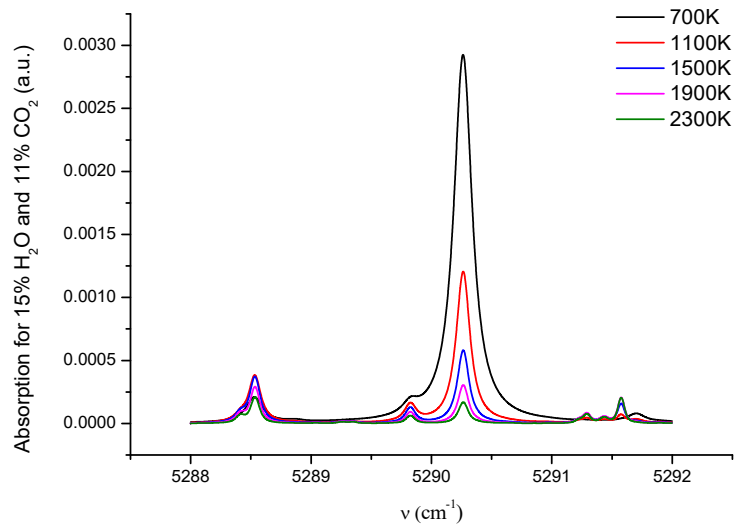

Figure 6: Simulated absorption spectra of $H_2O$ around 5290.264 $cm^{-1}$ (for temperature range from 300K to 2300K and 1 atm pressure) for pathlength 0.1 cm

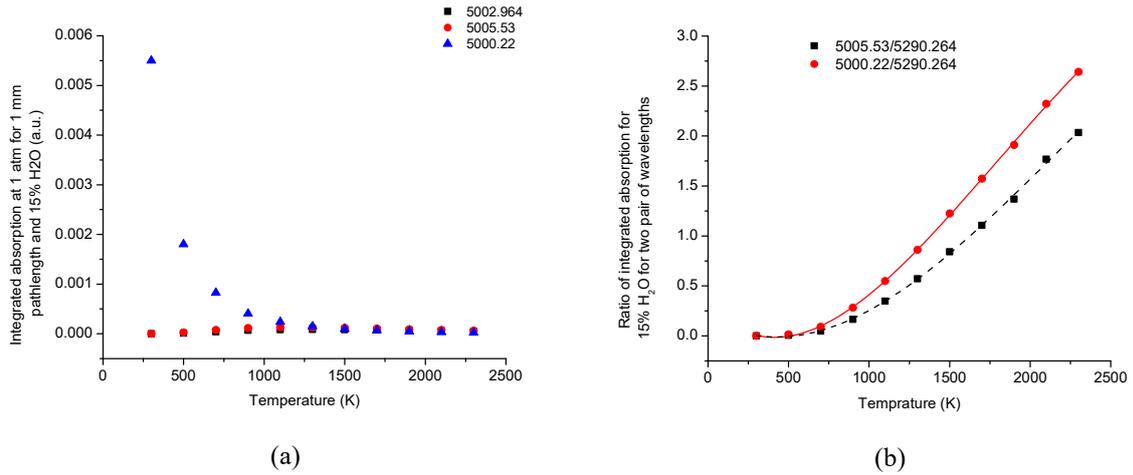

(a)                                                                 (b)

Figure 7: Variation of (a) integrated absorption for three wavenumbers 5290.264, 5005.53 and 5000.22 cm$^{-1}$,  (b) ratio of integrated absorption for two pairs for 15% $H_2O$ for 0.1 cm pathlength at 1 atm. Pressure and temperature range from 300K to 2300K.

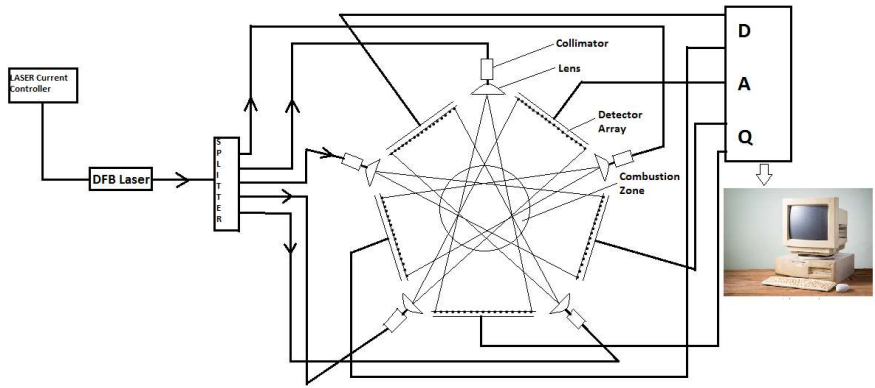

Figure 8(a): A Schematic diagram of a fan-beam tomographic system

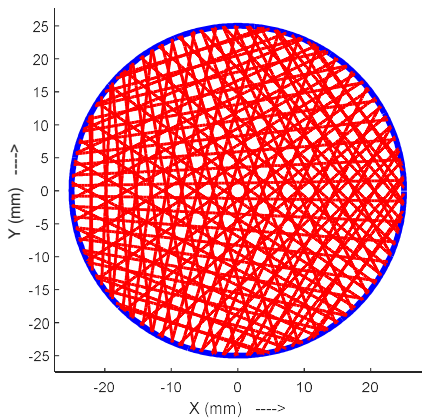

Figure 8(b): 5 fanbeam distribution of rays inside the circular object at center at (0, 0) and of diameter 50mm

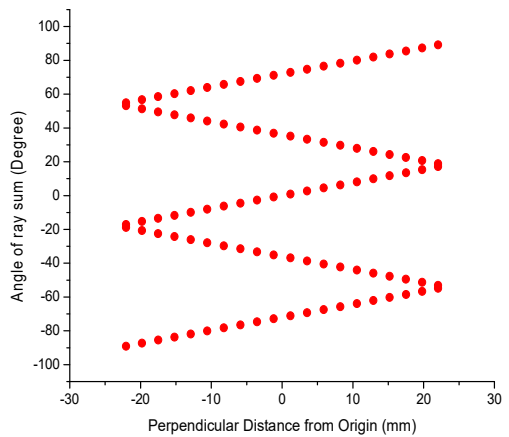

Figure 8(c) Sinogram plot of the beam arrangement as found in 8(a)

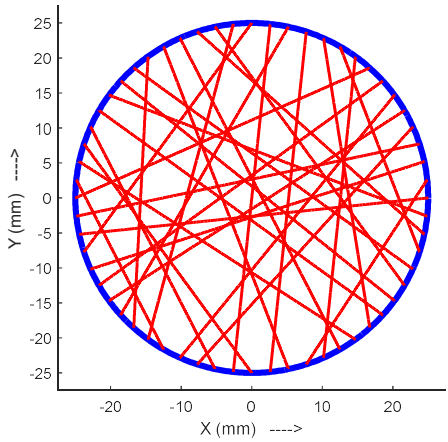 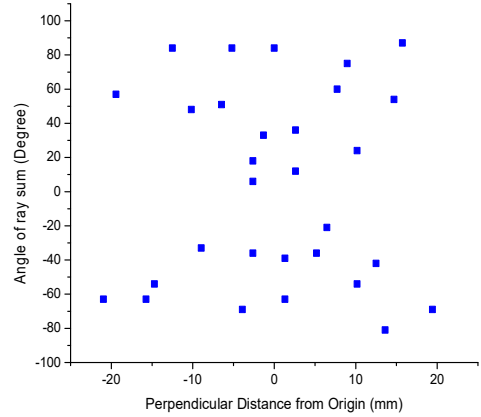

Figure 9(a): Discrete uneven distribution of rays inside the circular object at center at (0, 0) and of diameter 50mm

Figure 9(b): Sinogram plot of the beam arrangement as found in 9(a)

A schematic diagram of a tomographic system developed with fan-beam arrangement is shown in Figure 8(a). In this technique, the laser wavelength as well as intensity is modulated by a ramp current signal, typically in the range between 100Hz-10kHz. The distribution of ray lines within the object is shown in figure 8(b) and is also represented by its perpendicular distance from the centre of the object and its inclination angle with the x-axis, known as sinogram, as shown in figure 8(c).

The Distributed Feed-back (DFB) Laser has center wavelength $\sim 2\mu m$ and scanning range of 6 nm only by changing the driving current keeping the laser temperature constant. For the fan-beam geometry, its output is split into five channels by optical/fiber splitters. The outputs of such splitters are then collimated by suitable collimators and then diverged by plano-cylindrical lenses to obtain fan-beam geometry of incident rays and are launched through the absorbing medium. The transmitted rays from each fan beam are detected by an array of 20 Nos. photodetectors and the integrated absorbance $A_v$ for each ray is detected and normalized by the reference signal in non-absorbing channel (not shown in the figure 8(a)). As it is difficult to arrange so many photodetectors around the object in five fan-beam arrangements, the source and detectors are placed at distant places from the object. This in turn causes the beam to pass through the environment with low but variable moisture content resulting in corruption in the path-integral. This problem is solved by choosing the spectral lines for which the absorption values are comparable to the detection limit as discussed in wavelength selection criteria of section 3. From the quantities "$A_i$", the integrated absorption co-efficient $\alpha_j$ in j-th grid can be computed using some suitable algorithm discussed in section 5. If $\alpha_j$ is known for two transitions, $T_j$ can be inferred from suitable empirical equation (15), whence concentration $X_j$ can be found from using interpolation illustrated in section 3. As the number of grids in ROI is 1868, eliminating the masked region of 50x50 grid and the number of measurements are 5x20, i.e. 100, the tomography illustrated here is called "Limited Data Tomography".

For eliminating the absorption due to passage outside the object, a set of discrete rays were used to reproduce the same reconstructions using only 31 rays as compared to 100 in the earlier fan-beam case studies. The system hence should be much more undetermined than before and it's more challenging and difficult at the same time to extract out useful imaging information from such a system. However, Terzija et al. [26] showed if the representative points of the beam array in sinogram space are sparse and well covered in object space rather than closely densed as packets, then even a lesser number of beams may create reconstruction of good quality. The useful information as may be accessed from the system is more dependent on distribution of rays in sinogram space than their numbers. As a sample distribution, we try with a set of beam whose representative points are distributed like Figure 9(a) in Sinogram space. Figure 9(b) shows the sinogram distribution for the sparse discrete rays inside the object. In this scheme the laser beam is split by fiber splitter into 32 single collimated beams and 31 rays launched through the object in uneven distribution with 1 ray as reference and the photodetectors, placed just opposite the ray, capture the light to obtain the path integral of the absorption.

## 5. A Brief Review on Tomography-Algorithms:

The tomographic reconstruction can be represented as an inverse problem of the matrix equation (12). Considering $\boldsymbol{A} = [A_i], \boldsymbol{\alpha}_v = [a_j], \boldsymbol{L} = [L_{ij}]$, equation (12) reduces to,

$$\boldsymbol{\alpha}_v = \boldsymbol{L}^I \boldsymbol{A} \tag{19}$$

Since the right side of such matrix-equation often suffers from noise contamination from various sources in limited data tomography, a direct matrix-inversion often doesn't produce any meaningful solution (condition number of $\boldsymbol{L}$ can be very large making the system ill-conditioned). Also often the system is underdetermined algebraically making the problem ill-posed, precluding any possibility of direct matrix-inversion. A naïve solution, bearing the name linear back-projection [27], can however be written as,

$$\boldsymbol{\alpha}_v = \boldsymbol{L}^T \boldsymbol{A} \tag{20}$$

where $\boldsymbol{L}^T$ denotes transpose of $\boldsymbol{L}$. Though mathematically not quite accurate, this algorithm finds its widespread use in tomographic field due to its simplicity. However, for a better reconstruction, of course, this is not suitable and regularized solutions are to be used.

One possible unique least-square solution of the above matrix equation [28] can be written as:

$$\boldsymbol{\alpha}_v = \boldsymbol{L}^I \boldsymbol{A} = \sum_{i=1}^{n} \frac{v_i u_i^T A}{\sigma_i} \tag{21}$$

with $\boldsymbol{L}^I = \boldsymbol{V} \sum^I \boldsymbol{U}^T$ (called pseudoinverse) [29], where $\boldsymbol{L}$ can be decomposed as $\boldsymbol{L} = \boldsymbol{U} \sum \boldsymbol{V}^T$, satisfying,

$$\boldsymbol{U}^T \boldsymbol{U} = \boldsymbol{V}^T \boldsymbol{V} = \boldsymbol{I} \ ; \ \boldsymbol{U} = (\boldsymbol{u}_1, \boldsymbol{u}_2, \dots \boldsymbol{u}_n); \boldsymbol{V} = (\boldsymbol{v}_1, \boldsymbol{v}_2, \dots, \boldsymbol{v}_n); \sum = diag(\sigma_1, \sigma_2 \dots \sigma_n); \sigma_1 \geq \sigma_2 \geq .. \sigma_n$$

A regularized version of this solution can be written [29] as,

$$\boldsymbol{\alpha}_{v_{reg}} = \sum_{i=1}^{n} f_i \frac{v_i u_i^T A}{\sigma_i} \tag{22}$$

where, $f_i$ is a filter-factor dependent on the associated regularization technique. It can be showed that the famous regularization schemes named Tikhonov regularization [30] can be expressed in the above form if we take,

$$f_{i,tikhonov} = \frac{\sigma_i^2}{\sigma_i^2 + \lambda^2} \tag{23}$$

From the above relation it is clear that for the singular values much smaller than $\lambda$, the SVD co-efficients are gradually damped. If $\lambda$ is properly chosen, it can be shown that the norm of the covariance matrix of the regularized solution is always less than (often much less) that of unregularized one [29] (assuming a white-noise corrupted right side),

$$\|cov(x_\lambda)\|_2 < \|cov(x)\|_2 \tag{24}$$

hence, showing the regularized solutions to be less sensitive to noise-contamination.

An iterated version of the above regularization can be expressed [30] as,

$$x_{k+1} = x_k + \mu (A^T A + \Gamma^T \Gamma)^{-1} A^T (b - A x_k) \tag{25}$$

The regularization-problem can also be alternatively attacked by iterative techniques, like Conjugate Gradient, Landweber, Kaczmarz (commonly known as ART) etc. methods and their variants.

The Conjugate-Gradient scheme can be described as a minimization problem [31,32] of a quadratic function,

$$Q(\boldsymbol{\alpha}_v) = \frac{1}{2}\boldsymbol{\alpha}_v^T L \boldsymbol{\alpha}_v - A^T \boldsymbol{\alpha}_v \tag{26}$$

and a minimizer $\alpha_v^*$ of $Q(\boldsymbol{\alpha}_v)$ is found to be a possible solution of the matrix-equation above. Hence a line-search method with recurrence-relations is employed like,

$$\boldsymbol{\alpha}_{v,k+1} = \boldsymbol{\alpha}_{v,k} + \gamma_k \boldsymbol{p}_k \tag{27}$$

$$\boldsymbol{r}_{k+1} = \boldsymbol{r}_k - \gamma_k L \boldsymbol{p}_k \tag{28}$$

where,

$$\gamma_k = \frac{r_k^T r_k}{p_k^T P p_k} \tag{29}$$

$$\boldsymbol{p}_{k+1} = \boldsymbol{r}_{k+1} + \beta_k \boldsymbol{p}_k \tag{30}$$

where,
$$\beta_k = \frac{r_{k+1}^T r_{k+1}}{r_k^T r_k} \tag{31}$$

Here $r_k$ refers to $r(\boldsymbol{\alpha}_{v,k})$, with $r(\boldsymbol{\alpha}_v) = A - L\boldsymbol{\alpha}_v$.

Landweber [33] and Kaczmarz [34] iteration formula can be written respectively as,

$$\boldsymbol{\alpha}_{v,k+1} = \boldsymbol{\alpha}_{v,k} + \mu L^T (A - L\boldsymbol{\alpha}_{v,k}) \tag{32}$$

$$\boldsymbol{\alpha}_{v,k+1} = \boldsymbol{\alpha}_{v,k} + l_j \frac{A_j - l_j^T \boldsymbol{\alpha}_{v,k}}{l_j^T l_j} \tag{33}$$

Here $\mu$ is a gain-factor to control the convergence rate and $l_j$ is the j-th column-vector of matrix $L$. For a better response, a filtered version (e.g. median filter) of the above are used like,

$$\boldsymbol{\alpha}_{v,k+1} = f(\boldsymbol{\alpha}_{v,k} + \mu L^T (A - L\boldsymbol{\alpha}_{v,k})) \tag{34}$$

$$\boldsymbol{\alpha}_{v,k+1} = f\left(\boldsymbol{\alpha}_{v,k} + l_j \frac{A_j - l_j^T \boldsymbol{\alpha}_{v,k}}{l_j^T l_j}\right) \tag{35}$$

helping the reconstructions to add further noise-immunity.

## 6. Simulation and Results:

6.1. Uniform species-concentration and non-uniform temperature

A bimodal and a concentric unimodal hard-edged (top-hat) phantom for temperature distribution are assumed in the region of combustion for understanding the ability of different reconstruction algorithms for two pairs of wavelengths- first pair 5005.53 & 5003.3 cm$^{-1}$ and second pair 5000.22 & 5000.06 cm$^{-1}$. The modeled distribution of the temperature with constant concentration in terms of absorption coefficient can be displayed as colour map image [Figures 10(a) and 10(b)].

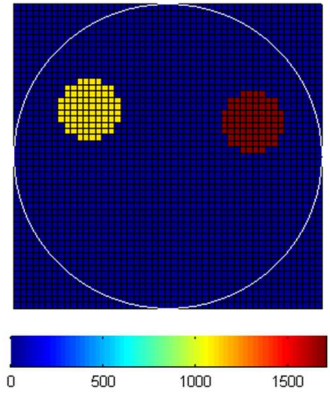 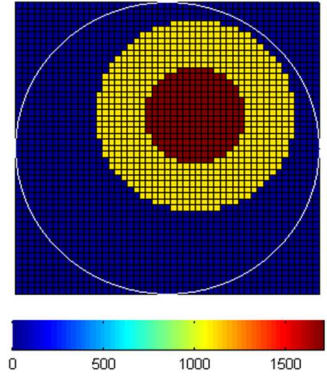

Figure 10(a): Bimodal top hat phantom having uniform 15% $H_2O$ and 11% $CO_2$ concentration with distribution of 500K background, hard edged two top hats at 1100 K and 1700 K.

Figure 10(b): Concentric top hat phantom having uniform 15% $H_2O$ and 11% $CO_2$ concentration with distribution of 500K background, two concentric top hats 1700K surrounded by 1100K.

A $50 \times 50$ grid system is used to discretize the ROI. The lowest integrated absorption signal for 15% $H_2O$ and 0.1 cm pathlength at wavenumber 5003.3 $cm^{-1}$ to be detected is $10^{-5}$ of the transmitted signal as found in figure 3. The white Gaussian noise is considered as one tenth of the absorption signal for 0.1 cm and background 2% $H_2O$ at 500K, which is of the order of $10^{-6}$ of the transmitted signal equivalent to $SNR{\sim}120 dB$. In the forward problem integrated absorptions of incident laser light at multiple paths have been calculated for two different wavelengths of a pair and the image of the distribution in terms of absorption has been reconstructed using different reconstruction techniques like filtered Landweber, filtered Tikhonov, Linear Back Projection (LBP) and Conjugate Gradient Least Square (CGLS). The ratio of the absorption values for two wavelengths of each pair for every grid has been calculated and the effective temperature of every grid is evaluated from ratio using empirical relation [equation (15)]. The temperature distribution is reconstructed and displayed in the colour map image.

Figure 11 shows a relative simulated distribution for bimodal hard-edged phantoms of 5mm diameter at (-13,7) and (13,5) at 1100 and 1700 K respectively using one pair of wavelengths (5005.53 and 5003.3 $cm^{-1}$) following four different reconstruction algorithms as stated above. The first two rows show the distribution of integrated absorption coefficient and the next two rows show ratio and temperature, respectively. Figure 12 shows the same for concentric hard-edged phantoms (1700 K of dia. 8 mm at center (3,4) surrounded by 1100 K of outside diameter 16 mm). For both the cases, concentrations of $H_2O$ and $CO_2$ are same all over the object as 15% and 11% respectively with Nitrogen as base gas. From the figure 11, it has been found that Tikhonov and Landweber show best results among the four. The maximum temperature attained by Tikhonov and Landweber is around 1700K, whereas the maximum temperature attained by LBP and CG are 1200K and 700K respectively. For LBP and CG, the temperature phantoms are spread out at the surrounding zones resulting in lower peak temperature than Landweber and Tikhonov. The same observation is true for concentric distribution of temperature as shown in Figure 12.

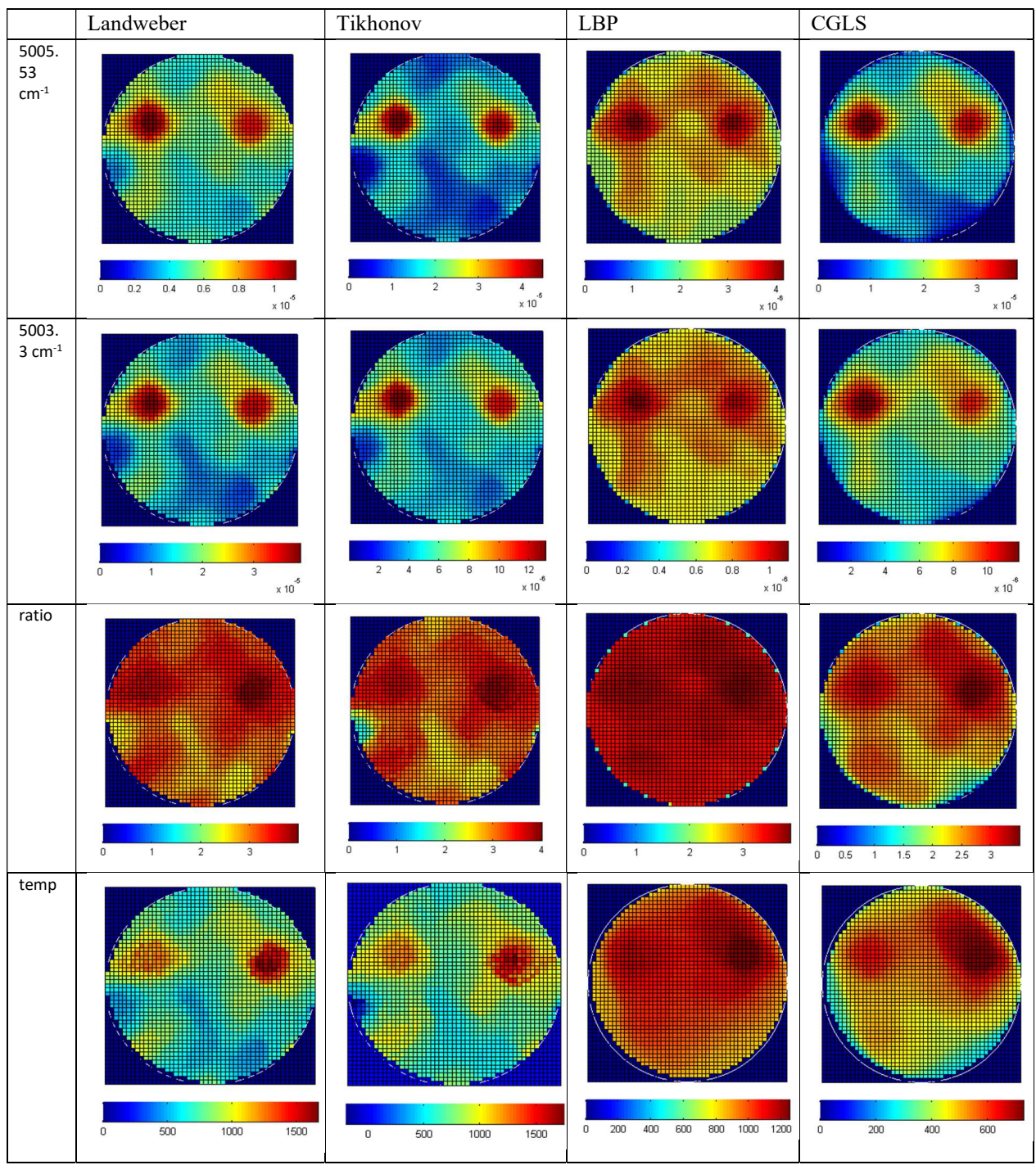

Figure 11. Reconstructions of distribution of absorption coefficients, ratio of absorption and temperature using the pair of spectral lines (5005.53, 5003.3) $cm^{-1}$ for bimodal top hat phantom

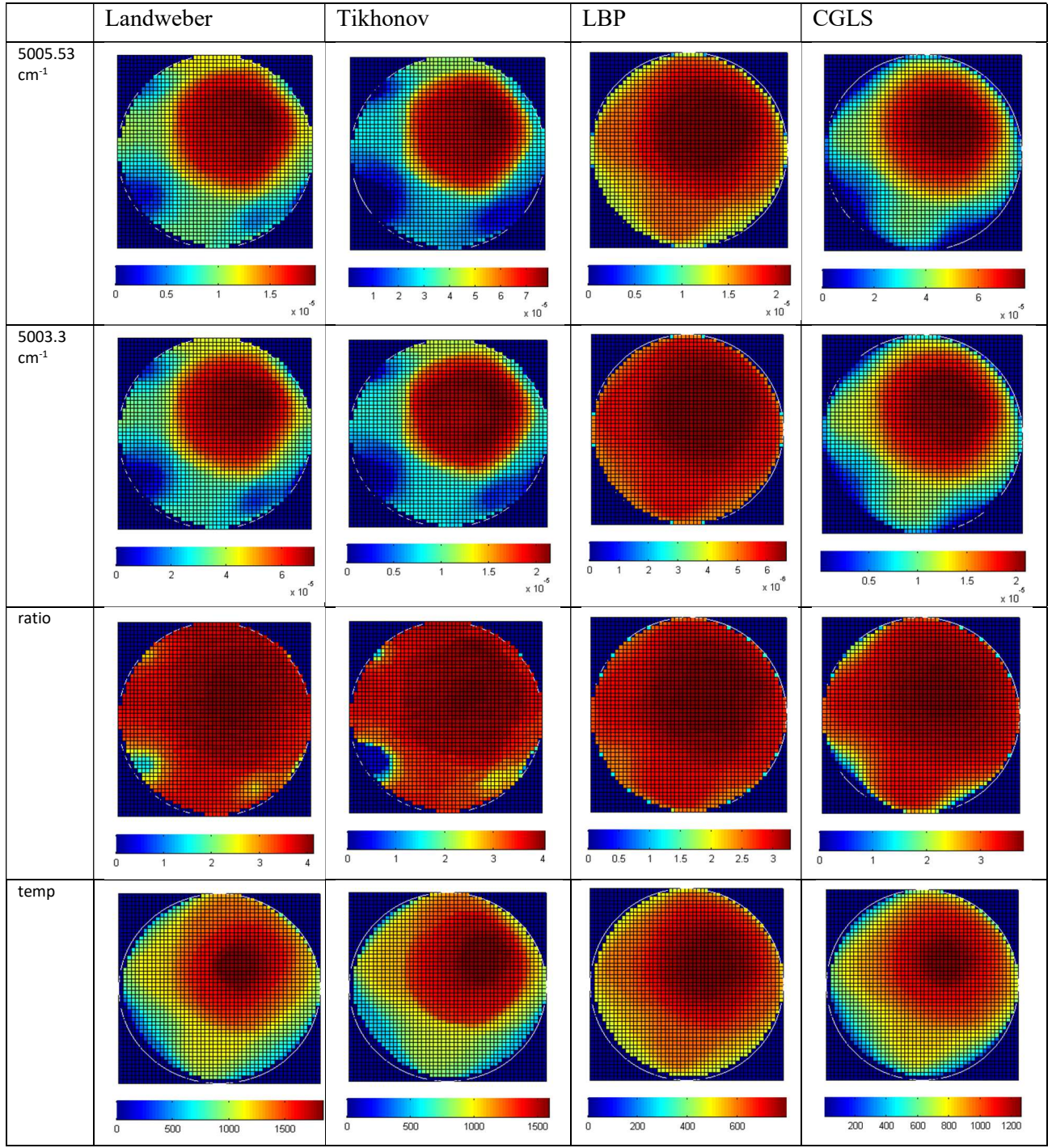

Figure 12: Reconstructions of absorption coefficients, ratio of absorption and temperature using the pair of spectral lines (5005.53, 5003.3) $cm^{-1}$ for concentric top hat temperature phantom

The similar reconstructions for separate and concentric top-hat phantoms were repeated for the other probable candidate, wavenumber pair 5000.22 and 5000.06 cm[-1] and are shown in figures 13 and 14 respectively. In this case also, reconstructed images formed by Tikhonov and Landweber algorithms represent much more distinct phantoms of about 1700K peak than the other two among four different algorithms. The reconstructed images for the bimodal top hat plumes using the later pair shows little poorer result than the images for earlier case, as the lower temperature plume attains a little lower value.

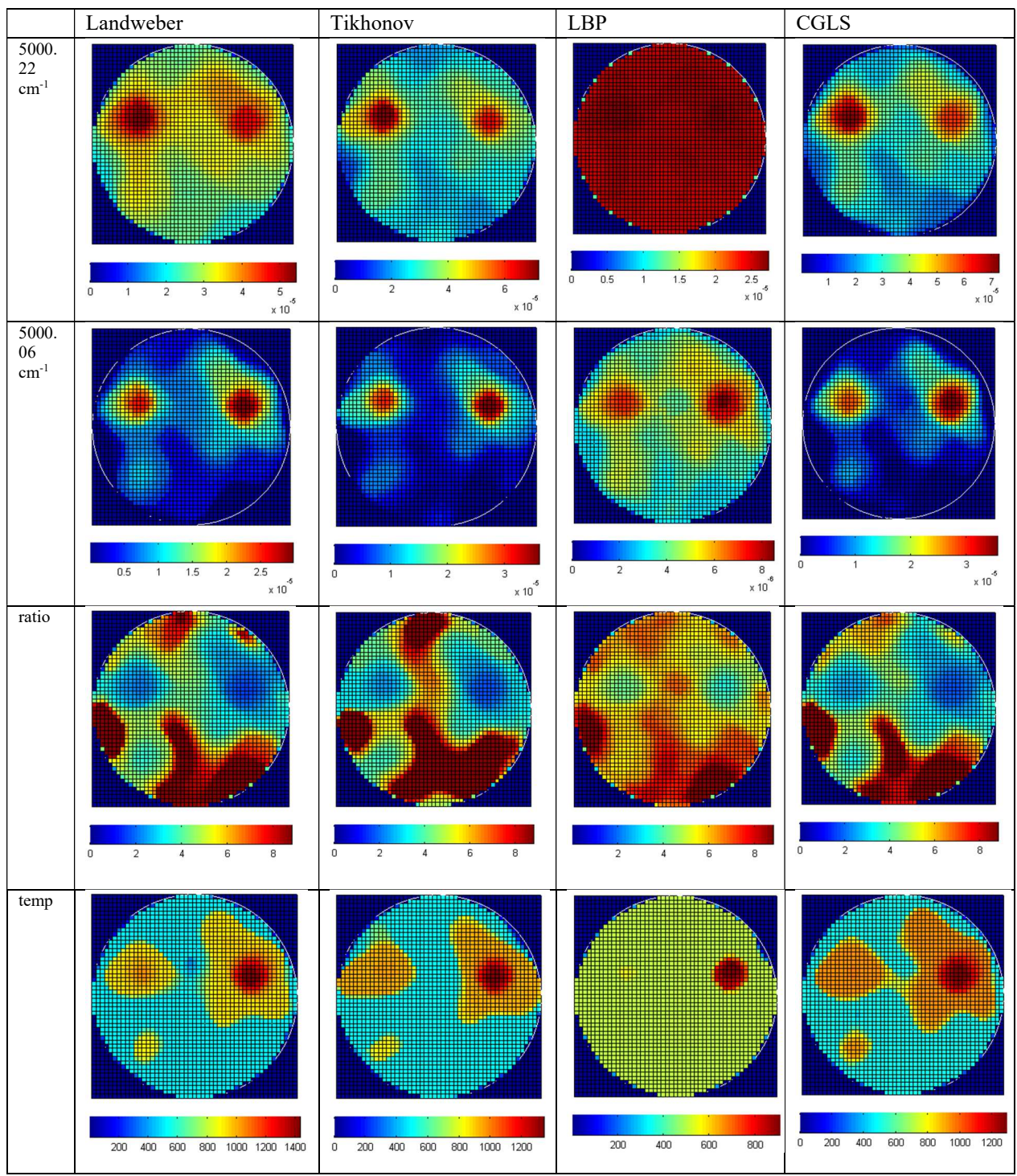

Figure 13: Similar reconstruction results as found in figure 11 for another pair of wavenumbers 5000.22 and 5000.06 cm$^{-1}$

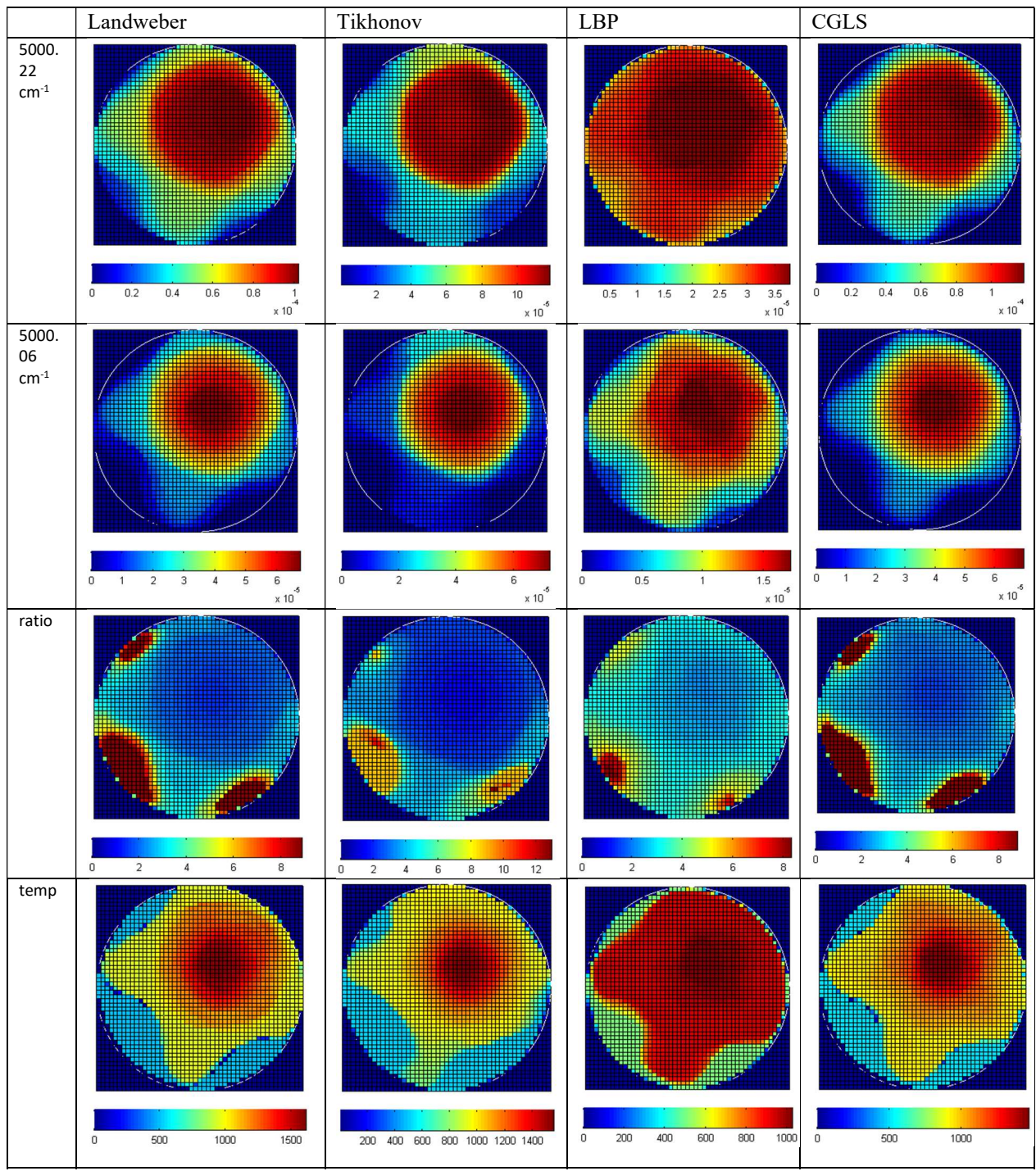

Figure 14: Similar reconstruction results as found in figure 11 for another pairs of wavelength 5000.22 and 5000.06 cm$^{-1}$

*6.2. Non-uniform species-concentration and non-uniform temperature:*

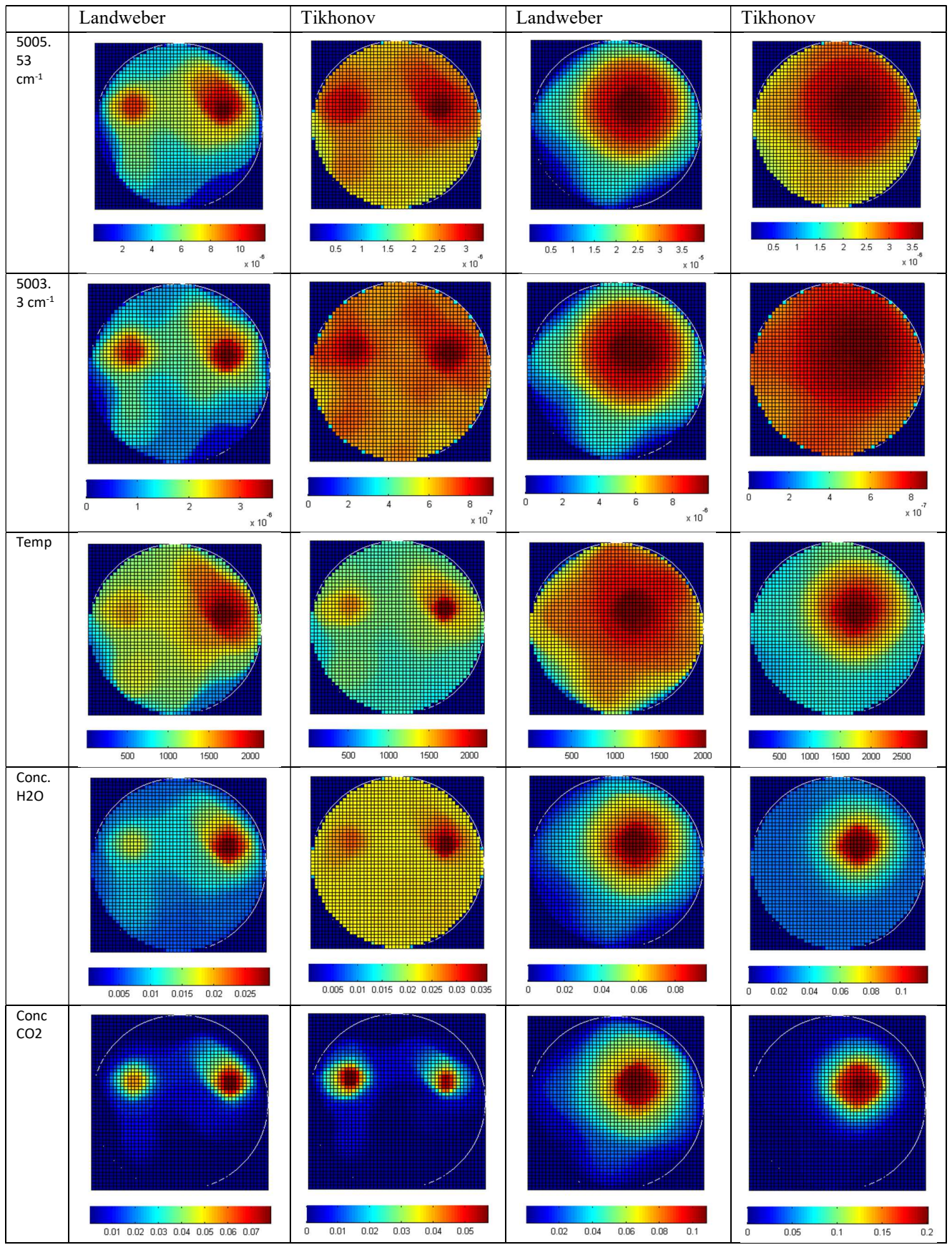

Figure 15: Reconstructions of variable concentration and temperature distributions (5005.53, 5003.3) $cm^{-1}$

Now, it is necessary to study the efficacy of this algorithm in reconstruction of images in which temperature and concentrations for $H_2O$ and $CO_2$ are both varying. As efficient combustion always generates maximum heat and combustion products, 1700K, 15% and 11% respectively in one plume and 1100K, 10% and 8% respectively in another plume are chosen. The integrated absorption values used for solving the forward problem are shown in table 3. The results so obtained are also presented in figure 15, using both Landweber and Tikhonov reconstruction. It is found that using Landweber reconstruction technique the images generated for temperature and concentrations of $H_2O$ and $CO_2$ are following nearly exact representation of two plumes in concentric cases, but those for two separate plumes are representative one - the concentration of $H_2O$ shows maximum 2.7% in place of 15% but that for $CO_2$ is 8% in place of 11%. This dilution is due to filtering effect of filtered Landweber which distributes the concentration throughout the surrounding. For concentric plumes, this effect is not as predominant as for bimodal plumes because for concentric plumes their respective diameters are 16 mm and 32 mm greater than that (10 mm) for bimodal plumes.

Table 3: Integrated absorption values for 0.1 cm length used for simulation derived from Spectraplot [35]

| Wavenumber (cm$^{-1}$) | | 15% H2O & 1700K | 15% H2O & 1100K | 10% H2O & 1100K | 2% H2O & 500K |
|---|---|---|---|---|---|
| H$_2$O | 5005.53 | 7.3713 E-05 | 8.3079 E-05 | 5.2207 E-05 | 1.4974 E-06 |
| H$_2$O | 5003.30 | 1.8437 E-05 | 2.2982 E-05 | 1.5363 E-05 | 5.9555 E-07 |
| H$_2$O | 5290.264 | 6.6641E-05 | 2.3863 E-04 | 1.5900 E-04 | 0.00024 |
| | | 11% CO2 & 1700K | 11% CO2 & 1100K | 8% CO2 & 1100K | 0.04% CO2 & 500K |
| CO$_2$ | 5004.37 | 2.8723E-06 | 1.2003 E-05 | 8.6303E-06 | 2.1200 E-07 |

The system is indeed highly underdetermined as most of the real-world tomographic problems with good temporal resolution. The convergence of the iterative algorithms is finally determined by the difference between actual value and derived value of integrated absorbance, $\varepsilon_{Int\_abs}$. To determine robustness of various algorithms for different phantoms, as a quality parameter, the r.m.s. and absolute value of relative error of temperature and concentration reconstructions are defined by equations (36) and (37), respectively.

$$\varepsilon_{rms}^T = \sqrt{\frac{\sum_{j=1}^{J}\left(\left((T_j - T_j')/T_j\right)^2\right)}{J}} \text{ and } \varepsilon_{abs}^T = \frac{\sum_{j=1}^{J}\left(|(T_j - T_j')|/T_j\right)}{J} \quad (36)$$

$$\varepsilon_{rms}^\chi = \sqrt{\frac{\sum_{j=1}^{J}\left(\left((\chi_j - \chi_j')/\chi_j\right)^2\right)}{J}} \text{ and } \varepsilon_{abs}^\chi = \frac{\sum_{j=1}^{J}\left(|(\chi_j - \chi_j')|/\chi_j\right)}{J} \quad (37)$$

where, $T_j$ (or $\chi_j$) and $T_j'$ (or $\chi_j'$) are exact and reconstructed temperature (or concentration) of $j$-th grid respectively. These values along with optimized iterative conditions are tabulated for filtered Landweber and Tikhonov algorithms in Table 4 for the best pair of $H_2O$ lines and both bimodal and concentric top hat distributions. To achieve the above reconstructions, regularization parameter and number of iterations chosen for optimized Landweber and Tikhonov solution have been shown in Table 4.

From Table 4 it is evident that both in terms of r.m.s. and absolute error as was defined in eqn. (36) and (37) filtered Tikhonov regularization produces better reconstruction than filtered Landweber in most cases except for $CO_2$. This is because the initial (non-vanishing) approximation of solution affects the reconstruction of image quality and Tikhonov regularization produces more accurate initial approximation than that of LBP, which is used in Landweber iteration, as depicted in the following equations:

$$x_{k+1} = x_k + \lambda A^T(b - Ax_k) \; [iterated\ Landweber] \quad (38)$$

$$x_{k+1} = x_k + \lambda(A^TA + \Gamma^T\Gamma)^{-1}A^T(b - Ax_k) \; [iterated\ Tikhonov] \quad (39)$$

In the first of above two equations, taking $x_0=0$, the initial non vanishing approximation $x_1$ for Landweber becomes $A^T b$, same as LBP. In the second equation, the approximation becomes $x_1=(A^T A + \Gamma^T \Gamma)^{-1} A^T b$ which is a better approximation than LBP. This is because this is a solution of the minimizer $\min_x \frac{1}{2}\|Ax - y\|^2+\|\Gamma x\|^2$, by which preference can be given to a particular solution of desirable properties. From the simulated reconstructions, it is evident that between Landweber and Tikhonov regularization, Tikhonov produces a better approximation than Landweber.

In comparison to separate bimodal cases, concentric cases present more accurate reconstruction in case of $H_2O$ but less accurate in case of $CO_2$. The decrease in absorption values for $CO_2$ with temperature results in higher error in $CO_2$ image reconstruction. The above results are also manifested in the reconstructed image shown in figure 14. The most noteworthy and interesting fact is that the r.m.s. errors of each kind of temperature-distribution is larger than that of $H_2O$ concentration. This is due to the fact that filtered iterative backprojection technique flattens the signature around the top-hat phantom, resulting in higher r.m.s. error.

Table 4: Presentation of overall r.m.s. and absolute mean error considering all the grids for temperature and concentration of $H_2O$ and $CO_2$ with iterative conditions selected for 5005.53 and 5003.3 cm$^{-1}$

| Errors after reconstruction | | Bimodal | | Concentric | | Iterative Conditions | Bimodal | | Concentric | |
|---|---|---|---|---|---|---|---|---|---|---|
| | | Landweber | Tikhonov | Landweber | Tikhonov | | Landweber | Tikhonov | Landweber | Tikhonov |
| Temperature | $\varepsilon_{rms}$ | 0.5611 | 0.5157 | 0.5015 | 0.4931 | $\lambda^*/N^*$(L) $\lambda^*/N^*/\alpha$(T) for $\nu_1$ | 0.00098/14 | 0.008/11/1 | 0.00098/14 | 0.00305/11/0.5 |
| | $\varepsilon_{abs}$ | 0.7328 | 0.7112 | 0.6787 | 0.6933 | $\varepsilon_{Int\_abs}$# | 9.2586E-05 | 0.00022092 | 0.00032885 | 0.0010453 |
| Conc. - $H_2O$ | $\varepsilon_{rms}$ | 0.5760 | 0.2647 | 0.4760 | 0.4093 | $\lambda^*/N^*$(L) $\lambda^*/N^*/\alpha$(T) for $\nu_2$ | 0.00087/14 | 0.008/6/1 | 0.00087/14 | 0.00305/6/0.5 |
| | $\varepsilon_{abs}$ | 0.7402 | 0.4621 | 0.6504 | 0.6120 | $\varepsilon_{Int\_abs}$# | 2.6376E-05 | 5.8867E-05 | 0.00010044 | 0.00028122 |
| Conc. – $CO_2$ | $\varepsilon_{rms}$ | 0.7484 | 0.8228 | 0.8183 | 1.6788 | $\lambda^*/N^*$(L) $\lambda^*/N^*/\alpha$(T) for $CO_2$ | 0.01/6 | 0.8/7/1 | 0.001/4 | 0.085/10/20 |
| | $\varepsilon_{abs}$ | 0.7803 | 0.7626 | 0.8553 | 1.0653 | $\varepsilon_{Int\_abs}$# | 1.8361E-06 | 8.7062E-07 | 9.2817E-05 | 0.00010604 |

* $\lambda$ and/or $\alpha$ for $\nu_1$ and $\nu_2$ : Regularization parameters for Landweber and Tikhonov;

N: No. of iterations

# $\varepsilon_{Int\_abs}$ : Absolute error after complete iterations on integrated absorption

Spectral line pair 5005.53 / 5290.264 cm$^{-1}$ is designated as pair #3. As pair #3 has higher sensitivities for temperature detection as discussed in section 2, the reconstruction using this pair will be used to check the reasonableness of using other neighbouring pairs for the same reconstruction. The same reconstructed images for temperature and concentration of $H_2O$ and $CO_2$ using pair #3 are shown in figure 16. From the images it is evident that the concentric images using filtered Landweber iterative technique show very distinct images but for others it is not true, though this pair has higher temperature sensitivity than other pairs. The errors after reconstruction for pair #3 are relatively less than pair #1 but the reconstruction for bimodal phantoms for $CO_2$ using both Tikhonov regularization and Landweber iteration produces larger error than pair #1, as evident from table 4 and 5. Since the improvement using pair #3 is not significant, it is wise to use the wavenumbers 5005.53 and 5003.3, possible to image temperature and concentration of $H_2O$ using a single laser source.

The above observation can be analyzed from the expression of relative errors. The relative r.m.s. error in $T$ is transmitted as a function of relative r.m.s. errors in $\alpha_1$ and $\alpha_2$ (using equaton (18)):

$$\varepsilon_{rms} \equiv \left(\delta T_j/T_j\right)_{rms} \approx \frac{k_B T_{rms}}{hc\Delta E''}\left(\frac{\delta R_j}{R_j}\right)_{rms} = \frac{\left(\left(\frac{\delta\alpha_{1j}}{\alpha_{1j}}\right)^2_{rms} - 2\left\langle\left(\frac{\delta\alpha_{1j}}{\alpha_{1j}}\right)\left(\frac{\delta\alpha_{2j}}{\alpha_{2j}}\right)\right\rangle + \left(\frac{\delta\alpha_{2j}}{\alpha_{2j}}\right)^2_{rms}\right)^{\frac{1}{2}}}{\left(hc\Delta E''/k_B T_{rms}\right)} = \frac{e}{\xi} \quad (38)$$

The relative r.m.s. error of temperature decreases with relative sensitivity of the transition pairs. From the ratio vs. temperature curve, it is clear that the sensitivity gradually decreases towards high temperature for both the transition pair #1 and #3.

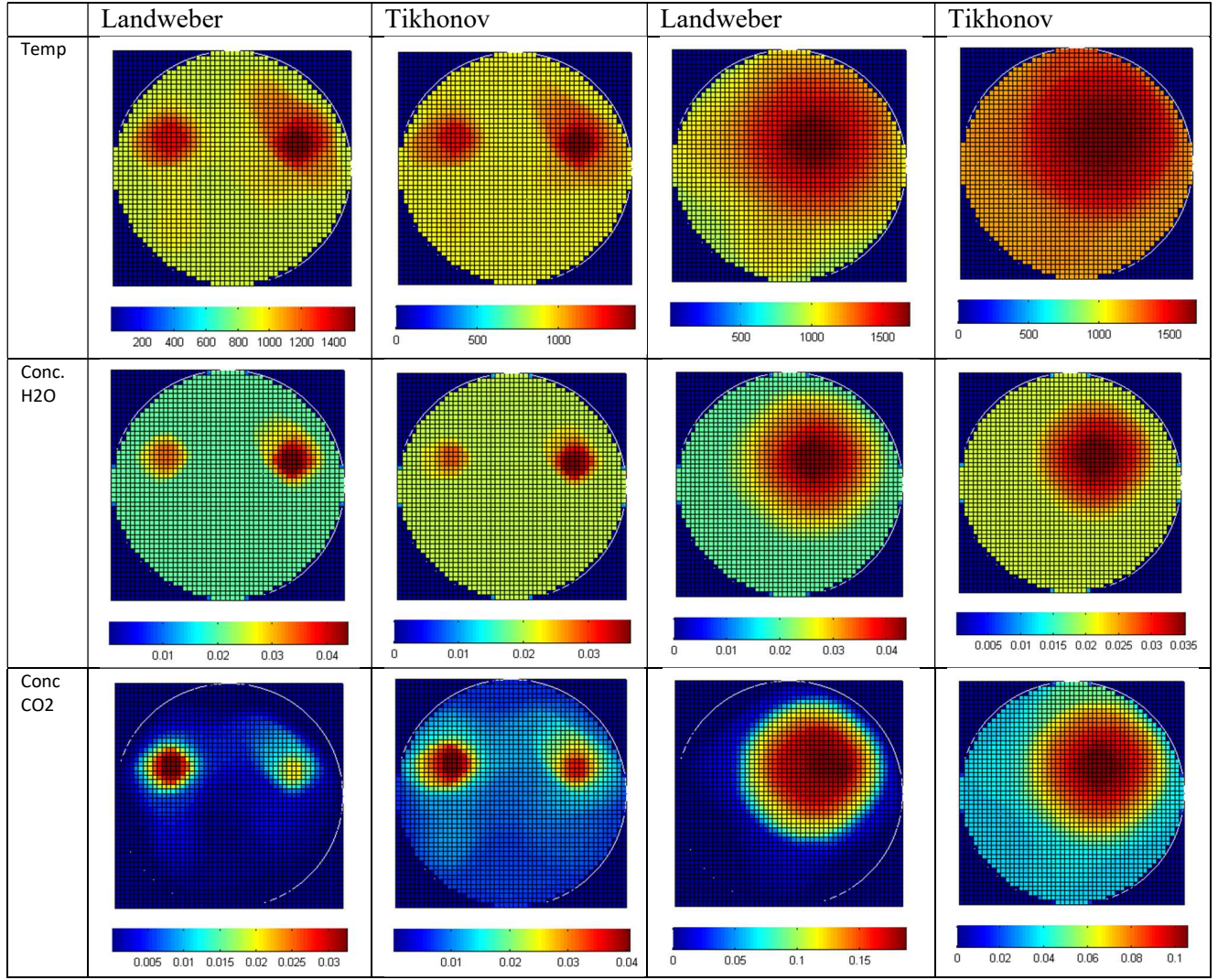

Figure 16: Reconstructions of variable concentration and temperature distributions (5005.53, 5290.264) $cm^{-1}$

Table 5: Presentation of overall r.m.s. and absolute mean error considering all the grids for temperature and concentration of $H_2O$ and $CO_2$ with iterative conditions selected for 5005.53 and 5290.264 cm$^{-1}$

| Errors after reconstruction | | Bimodal | | Concentric | |
|---|---|---|---|---|---|
| | | Landweber | Tikhonov | Landweber | Tikhonov |
| Temperature | $\varepsilon_{rms}$ | 0.4507 | 0.4445 | 0.5560 | 0.6024 |
| | $\varepsilon_{abs}$ | 0.6623 | 0.6584 | 0.7314 | 0.7633 |
| Conc. - $H_2O$ | $\varepsilon_{rms}$ | 0.2265 | 0.2035 | 0.4841 | 0.3399 |
| | $\varepsilon_{abs}$ | 0.3009 | 0.2707 | 0.5545 | 0.4295 |
| Conc. – $CO_2$ | $\varepsilon_{rms}$ | 1.0577 | 0.9876 | 1.0646 | 0.9673 |
| | $\varepsilon_{abs}$ | 0.8426 | 0.9758 | 0.9267 | 0.9752 |

Another important observation is that the integrated absorption for 5290.264 cm$^{-1}$ follows inverse relation with temperature. In the reconstructed image, the plumes region has less absorption than its surrounding region and the absorption level for the plumes is a small fraction of the total absorption. Hence during reconstruction, the contrast is lost in the plume region resulting in ingress of large error in temperature reconstruction as obvious from figure 16. Therefore, the benefit of higher sensitivity for pair #3 would be lost in temperature and concentration reconstruction process. This would rule out the benefit of using two lasers for achieving better sensitivity.

It is evident that if one set of sparse irregularly distributed beams is well covered in Sinogram (as compared to 9(b) where the points are not well scattered but linearly densed), then the set can be a good candidate to image temperature and concentration in the cases we studied before. Using this set of beam, the reconstructed images are displayed in Figure 17 (using the pair 5005.53/5003.3 cm$^{-1}$) which shows such an array of beam is highly prospective to develop a real-world tomographic system. The efficacy of the above unevenly distributed single beam arrangement is judged by the r.m.s. and absolute error as shown in table 6. One noteworthy fact is that many of the errors of reconstruction are less than the errors produced by fan-beam reconstructions despite reduction of beam number. This observation is in conformity with the study made by Terzija et al. as we explained before.

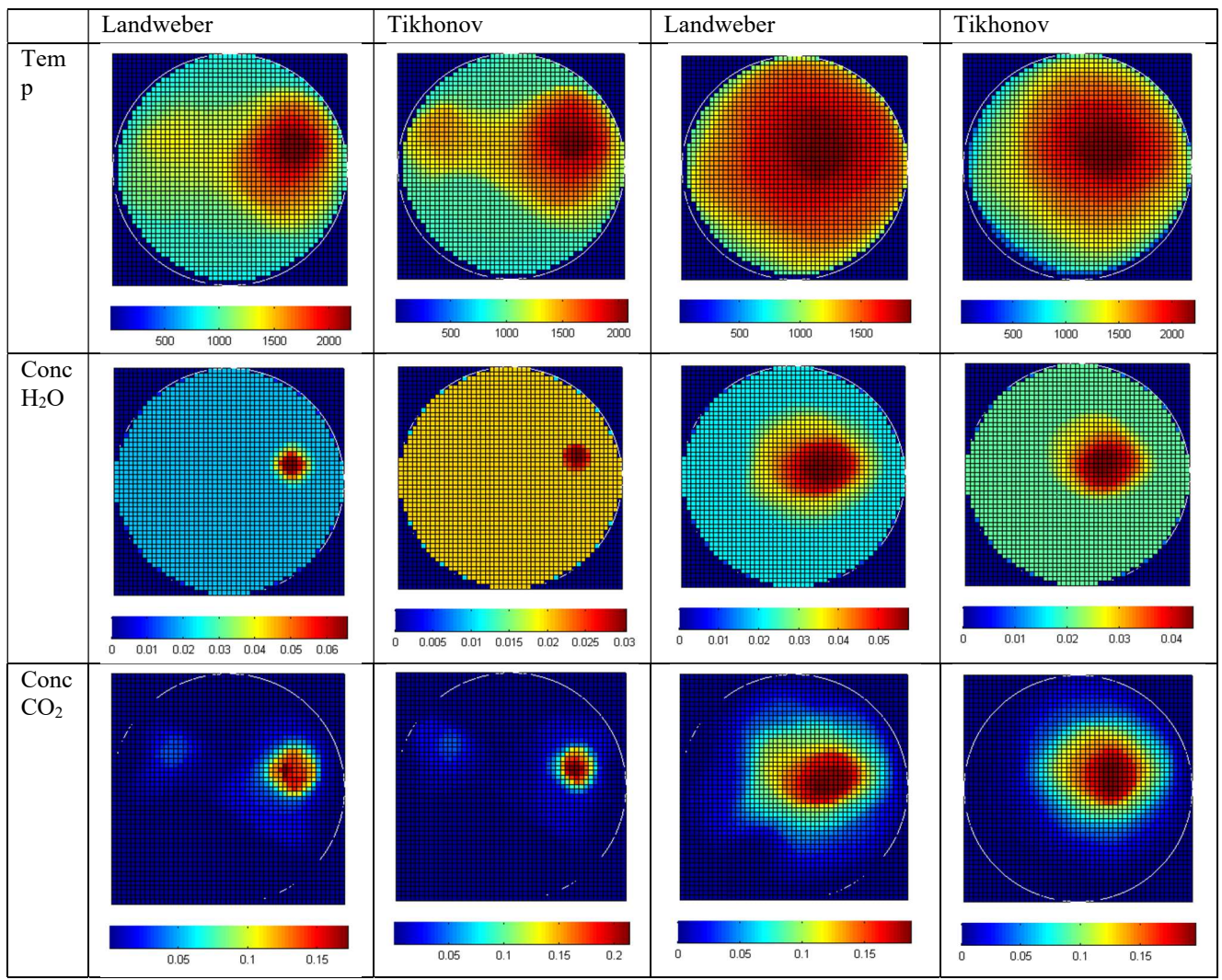

Figure 17: Reconstruction of Temperature and concentrations of H$_2$O and CO$_2$ in case of discrete unevenly distributed beams using Filtered Landweber and Tikhonov algorithm for both bimodal and concentric plumes in an object for the pair 5005.53 and 5003.3 cm$^{-1}$

Table 6 : Ultimate r.m.s. and absolute error using Landweber and Tikhonov algorithm for Bimodal and Concentric cases in case of discrete unevenly distributed beams

| Errors after reconstruction | | Bimodal | | Concentric | |
|---|---|---|---|---|---|
| | | Landweber | Tikhonov | Landweber | Tikhonov |
| Temperature | $\varepsilon_{rms}$ | 0.5537 | 0.5521 | 0.4984 | 0.4899 |
| | $\varepsilon_{abs}$ | 0.7341 | 0.7324 | 0.6799 | 0.6794 |
| Conc. - $H_2O$ | $\varepsilon_{rms}$ | 0.2250 | 0.2392 | 0.4652 | 0.5132 |
| | $\varepsilon_{abs}$ | 0.2598 | 0.2648 | 0.5548 | 0.5814 |
| Conc. – $CO_2$ | $\varepsilon_{rms}$ | 1.1690 | 0.8809 | 0.9928 | 0.8341 |
| | $\varepsilon_{abs}$ | 0.8775 | 0.8427 | 0.8680 | 0.8496 |

## 7. Conclusion

The plausibility of reconstruction of the temperature and concentration distribution of $H_2O$ and $CO_2$ by tomographic process, utilizing three very closely lying spectral lines of $H_2O$ and $CO_2$ (which can be probed by a single narrowband laser diode) with center wavelength around $2\mu m$, can be justified clearly by the quality of reconstruction. Based on the wavenumber selection criteria explained in section 3, three pairs of transition lines of $H_2O$ were identified for temperature and concentration tomography such as pair #1: 5005.53 & 5003.3 $cm^{-1}$; pair #2: 5000.22 & 5000.06 $cm^{-1}$; and pair #3: 5005.53 & 5290.264 $cm^{-1}$. Pair #1 and pair #2 can be scanned by a single narrow band distributed feed-back (DFB) laser. To assess the quality of reconstruction, the same reconstructions using three pairs of $H_2O$ lines and one $CO_2$ line (5004.37 $cm^{-1}$) were carried out using two different phantom distributions (bimodal and concentric) and introducing white noise of 120 dB in projection data. Initially four reconstruction algorithms, i.e. Landweber, Tikhonov, LBP and CGLS were tried, but on the basis of performance of reconstruction Landweber and Tikhonov were taken for further reconstruction study. The quality of reconstruction was also studied using an irregular and sparse 31 beam system. The quality of reconstruction of temperature and various concentrations for pair #1 is the best among all three; in terms of the reconstructed image. This shows that the triplet of lines which has been used here is one of the best possible triplets to reconstruct the temperature and concentration distribution of $H_2O$ and $CO_2$, providing the privilege of using only one narrowband laser diode. It is also clear that only high sensitivity of ratio of integrated absorption with temperature would not help in temperature and concentration reconstruction process.

**Data Availability:**

The data that support the findings of this study are openly available in https://hitran.org/ [24] and https://www.spectraplot.com/ [35].